\documentclass[pra, aps, amsmath, amssymb, superscriptaddress, twocolumn]{revtex4-2}
\usepackage{subcaption}
\usepackage{graphicx} 
\usepackage{appendix}
\usepackage{amsthm}
\usepackage{amsfonts}
\usepackage{mathrsfs}
\usepackage{braket}
\usepackage{physics}
\usepackage{derivative}
\usepackage{xcolor}
\usepackage[font={footnotesize}]{caption}
\usepackage[colorlinks]{hyperref}
\usepackage{orcidlink}

\hypersetup{
  colorlinks=true,
  linkcolor=blue,
  urlcolor=blue,
  citecolor=blue,
  linktocpage=true
}
\newcommand{\graph}{\mathcal{G}}

\newcommand{\kryldim}{K}
\newcommand{\Krylket}[1]{\ket{K_{#1}}} 
\newcommand{\Krylbra}[1]{\bra{K_{#1}}} 

\newcommand{\comploftime}[1]{\mathcal{C}(#1)} 

\newcommand{\infavcompl}{\bar{\mathcal{C}}} 

\newcommand{\hamiltonian}{\hat{H}} 
\newcommand{\hilbert}{\mathcal{H}} 
\newcommand{\quasigauge}{\hat{\Lambda}_{\Vec{\varphi}}}

\newcommand{\dimension}{D}

\newtheoremstyle{mythmstyle}
  {}   
  {}   
  {\itshape} 
  {}       
  {\bfseries} 
  {}      
  {\newline}      
  {}       

\theoremstyle{mythmstyle}

\newtheorem{definition}{Def}

\newtheorem*{observation*}{\underline{Obs}}

\newtheorem{theorem}{Theorem}

\begin{document}

\begin{abstract}
Krylov spread complexity quantifies how extensively a quantum system explores its Hilbert space during time evolution. 
We identify two classes of unitary transformations that preserve the complexity of the associated dynamics, and relate this property to the invariance of the energy distribution of the initial state and of its associated Hamiltonian moments.
These transformations are either symmetry operator of the Hamiltonian, or local $U(1)$ transformations of which the initial state is an eigenstate. The latter naturally arise as quasi-gauge transformations in chiral continuous-time quantum walks, to which the invariance properties are applied. 
We show that chiral phases in circulant Hamiltonians generically lift spectral degeneracies, leading to a discontinuous change in the long-time-averaged complexity.
For a localized initial states and a nondegenerate spectrum, complexity is $\infavcompl=(\dimension-1)/2$, where $\dimension$ is the Hilbert-space dimension, independently of the degree of connectivity. A non-circulant quantum router illustrates the additional role of phase-dependent spectral weights. These examples show that Krylov complexity can change even when probability time-reversal symmetry is preserved, linking its response to chirality to spectral changes rather than directly to transport asymmetry.
%
\end{abstract}

\title{Symmetry and quasi-gauge invariance of Krylov complexity}

\author{Edoardo Ferretti\,\orcidlink{0009-0004-0790-3349}}
\email{edoardo.ferretti@physik.uni-freiburg.de}
\affiliation{Dipartimento di Fisica "Aldo Pontremoli", Università di Milano, I-20133 Milano, Italy}
\affiliation{Institute of Physics, University of Freiburg, D-79104 Freiburg, Germany}
\author{Matteo G. A. Paris\,\orcidlink{0000-0001-7523-7289}}
\email{matteo.paris@unimi.it}
\affiliation{Dipartimento di Fisica "Aldo Pontremoli", Università di Milano, I-20133 Milano, Italy}
\author{Claudia Benedetti\,\orcidlink{0000-0002-8112-4907}}
\email{claudia.benedetti@unimi.it}
\affiliation{Dipartimento di Fisica "Aldo Pontremoli", Università di Milano, I-20133 Milano, Italy}

\date{\today}

\maketitle

\section{Introduction}
The notion of complexity in quantum mechanics heavily depends on the circumstances of what is being characterized: depending on the context, complexity fragments into different definitions, often difficult to reconcile. For a single system, statistical complexity may be a convenient choice \cite{lopez1995statistical,Tang_2025,tang2026phase}, while in the context of many-body physics a system is labeled as complex if a significant portion of the overall Hilbert space is explored: such a property can be measured by operator size \cite{Roberts_2015, Roberts_2018}. On the other hand, the advancements in quantum computing platforms demand efficiency, pushing for a resource-wise notion of complexity, such the Nielsen gate complexity \cite{nielsen2005geometricapproachquantumcircuit}. A more universal notion of complexity should be able to encompass or bridge between such diverse definition, and a good candidate for this role is \textit{Krylov Complexity} \cite{PhysRevX.9.041017}. 
\par
This definition of complexity descends from Krylov reduction, a computational method widespread across linear algebra. It quantifies how much the time-evolved state spreads over the Krylov basis generated by the repeated action of the dynamical generator on the initial state, and it can be computed efficiently via the Lanczos algorithm \cite{NANDY20251,Lanczos:1950zz}. 
As all quantum systems are linear, this method can be universally applied and thus provides a robust definition of complexity. For this reason, in recent years Krylov complexity has attracted much attention from different directions of research. Its formulation for operators, for example, is useful as a probe for discriminating integral systems from chaotic ones \cite{Rabinovici_2021, Balasubramanian_2022, Rabinovici_2022, Espa_ol_2023, balasubramanian2023quantumchaosintegrabilitylate, Craps_2025, Alishahiha_2025, Baggioli_2025}. From the quantum computation area, Krylov complexity of a state has been proved to be related to the upper-bound of Nielsen complexity, opening the possibility to a geometric interpretation \cite{dowling2006geometryquantumcomputation,PhysRevLett.132.160402,spd2-l836,lv2023buildingkrylovcomplexitycircuit, Seetharaman_2025}. Complexity by itself also showcases interesting properties, such an emergent monotonicity law analogous to the second law of thermodynamics \cite{balasubramanian2025variationsthemekrylov}. Furthermore, the Krylov method provides a natural mapping between any dynamics and a one-dimensional hopping model, the Krylov chain, effectively described as a continuous-time quantum walk (CTQW) \cite{Farhi_1998, Kempe_2003, M_lken_2011, Kadian:2021dej, Jeevanesan_2024}. 
\par
Since Krylov complexity is defined relative to a choice of initial state and Hamiltonian, a first natural question is what it is actually sensitive to: which features of the dynamics does it capture, and which does it discard? Identifying maps between different dynamics that leave the associated complexity invariant is a step in such direction, and constitutes the main topic of this work. We investigate the invariance properties of Krylov complexity for isolated, undriven systems, and identify two classes of transformations that leave it, together with the whole distribution on the Krylov chain, unchanged. These are symmetry operators of the Hamiltonian (Theorem \ref{thm:complgeneralinv}) and unitary transformations for which the initial state is an eigenvector (Theorem \ref{thm:complinv}).
The former generalizes the known time-translation invariance of complexity \cite{PhysRevD.109.L081701} to arbitrary symmetry operators of the Hamiltonian, and to arbitrary moments on the Krylov chain. The latter finds a natural realization in \textit{quasi-gauge} transformations, well known and used in the context of  chiral continuous-time quantum walks  \cite{Zimbor_s_2013, PhysRevA.93.042302, Frigerio_2021, Turner_2021}, a class of CTQWs characterized by complex hopping amplitudes. The associated phases modify the interference between different propagation paths and can substantially affect the walker dynamics, enabling unique phenomena like directional transport, enhanced state-transfer efficiencies and improved performance in quantum-walk search algorithms \cite{ Wong2015,frigerio22, Yu_2023,Annoni_2024,finocchiaro25,ragazzi25,Benedetti_2026}. 

Therefore, chiral quantum walks provide a particularly convenient application of the general framework developed in this work. 
Using the newly established invariance theorems, together with analytical arguments and numerical computations, we investigate the effect of chirality on long-time-averaged Krylov complexity.
For circulant Hamiltonians, real and chiral walks display qualitatively different complexity scaling. For real Hamiltonians, the complexity depends on the graph topology through the degeneracy structure of the spectrum, whereas generic chiral Hamiltonians attain a universal connectivity-independent value. We also analyze a minimal quantum router, showing that the relationship between complexity, spectral delocalization and graph topology can be more subtle in non-circulant systems. 
These results suggest  that Krylov complexity is primarily sensitive to the spectral consequences of chirality,  such as degeneracy lifting, rather than directly to probability time-reversal-symmetry breaking, which is often associated with chiral dynamics.
\par
The paper is organized as follows. In Sec. \ref{sec:degeneracy} we review the Krylov reduction method, and define complexity. Since the Hamiltonias studied are in general degenerate, we carefully review the subtle intertwining between degeneracy and the construction of the Krylov basis, using the basis of representatives \cite{Rabinovici_2022, rabinovici2025krylovcomplexity} as a fundamental tool. Sec. \ref{sec:qwchiral} is dedicated to the introduction of chiral quantum walks and of quasi-gauge transformations. In Sec. \ref{sec:invariance} and \ref{sec:invprop} we establish the invariance of complexity under symmetry and quasi-gauge transformations, respectively. In Sec. \ref{sec:cctqw} we apply these results to compute complexity of chiral quantum walks on different topologies, to test its sensitivity to chirality. Sec. \ref{sec:discussion} closes the paper.

\section{Krylov reduction in the presence of degeneracy} \label{sec:degeneracy}

Consider an isolated quantum system associated with a Hilbert space $\hilbert$, with finite dimension $\dimension$. Initially, the system is found in a state $\ket{\psi_0}$, whose time evolution generated by the time-independent Hamiltonian $\hamiltonian$ is given by
\begin{equation} \label{eq:schrev}
    \ket{\psi (t)}=e^{-i\hamiltonian t}\ket{\psi_0} = \sum_{n=0}^{\infty} \frac{(-it)^n}{n!}\ket{\psi_n},
\end{equation} 
with $\ket{\psi_n} = \hamiltonian^n\ket{\psi_0}$. Since the Hilbert space is finite-dimensional, all terms in (\ref{eq:schrev}) are linearly dependent on (at most) the first $\dimension$ ones. Therefore, once orthonormalized, these constitute a basis of the minimal subspace explored by the dynamics, called the \textit{Krylov subspace} \cite{NANDY20251}. The most common procedure adopted for such orthonormalization is the Lanczos algorithm \cite{Lanczos:1950zz}, defined by the iterative step:
\begin{align}
    \ket{A_{n+1}} &:= \left ( \hamiltonian - a_n \right ) \Krylket{n} - b_n \Krylket{n-1}, \\
    \Krylket{n+1} &:= (b_{n+1})^{-1}\ket{A_{n+1}}.
\end{align}
The Lanczos coefficients are defined as $a_n\,:=\,\Krylbra{n}\hamiltonian\Krylket{n}$, $b_n^2 := \braket{A_n}$, with $b_0=0$ and $\Krylket{0}:=\ket{\psi_0}$. The algorithm stops at $n+1=\kryldim\leq \dimension$, when $b_{n+1}=0$. The Lanczos coefficients can be related through polynomial relations to the moments of the energy distribution of the initial state. Expressed in the Krylov basis, the Hamiltonian is represented by a tridiagonal matrix, with $\{a_n\}$ on the diagonal and $\{b_n\}$  the off-diagonals elements. 
Krylov vectors $\Krylket{n}$ can be represented as the result of acting on the initial state with polynomial functions of the Hamiltonian, namely $\Krylket{n} = P_n(\hamiltonian)\ket{\psi_0}$, with: 
\begin{equation} \label{eq:polclosedform}
    P_n(\hamiltonian) = \frac{1}{\prod_{i=1}^n b_i}  \det (\hamiltonian I_n - h_n),
\end{equation}
where $I_n$ is the $n\times n$ identity matrix and $h_n$ is a $n\times n$ tridiagonal matrix of Lanczos coefficients, representing $\hamiltonian$ in the Krylov basis \cite{MUCK2022115948,balasubramanian2025variationsthemekrylov}. For example, the first polynomials read
\begin{align}
    \begin{split}
        P_0(\hamiltonian) &:= \mathbb{I}, \\ 
        P_1(\hamiltonian) &= \frac{1}{b_1}\left ( \hamiltonian - a_0 \right ), \\
        P_2(\hamiltonian) &= \frac{1}{b_1b_2} \left [ \hamiltonian^2 - (a_0+a_1)\hamiltonian +a_0a_1 -b_1^2 \right ].
    \end{split}
\end{align}

The definition of the Krylov basis lets us map the dynamics onto a 1D lattice, whose sites are the Krylov states, with on-site energies given by $\{a_n\}$ and transition amplitudes between adjacent sites given by $\{b_n\}$. Therefore, studying the dynamics is equivalent to studying the distribution of the evolved state on such a chain, through the moments of the position operator 
\begin{equation}
    \hat{K} = \sum_{n=0}^{\kryldim -1 } n\Krylket{n}\Krylbra{n}.
\end{equation}
The first moment, also known in literature as \textit{Krylov spread complexity} is defined as
\begin{equation} \label{eq:compldef}
    \comploftime{t} := \sum_{n=0}^{\kryldim-1} n\,|\Krylbra{n}\ket{\psi(t)}|^2 = \bra{\psi(t)}\hat{K}\ket{\psi(t)}.
\end{equation}
Analogously, it is possible to define higher moments of $\hat{K}$, such as the \textit{Krylov variance} 
\begin{equation} \label{eq:krylvariance}
    \Delta^2K(t) = \bra{\psi(t)}\hat{K}^2\ket{\psi(t)}-\left ( \bra{\psi(t)}\hat{K}\ket{\psi(t)} \right )^2.
\end{equation}
By inserting two resolutions of the identity in the energy basis in the definition of complexity, we obtain
\begin{align} \label{eq:decomposedcompl}
    \begin{split}
        \comploftime{t} = \!\sum_{n,j,k}& n\!\braket{K_n}{E_j}\!\braket{E_j}{\psi_0} \\ &\times\!\braket{\psi_0}{E_k}\!\braket{E_k}{K_n}e^{-i(E_j-E_k)t}.
    \end{split}
\end{align}
Equation (\ref{eq:decomposedcompl}) highlights the oscillatory nature of complexity for isolated systems. For this reason, the long-time average of complexity, defined as:
\begin{equation} \label{eq:infavcompl}
    \infavcompl = \lim_{T\to \infty} \frac{1}{T}\int_0^T dt\text{ }\comploftime{t},
\end{equation}
is adopted here as a measure of how much the dynamics explores the Krylov subspace. Assuming the spectrum is in general degenerate, it holds that 
\begin{equation}
    \lim_{T\to \infty}\frac{1}{T}\int_0^T dt\text{ }e^{-i(E_j-E_k)t} = \delta_{E_jE_k},
\end{equation}
meaning that the limit of the integral is non zero whenever $\ket{E_j}$ and $\ket{E_k}$ belong to the same eigenspace of $\hamiltonian$. 
Let us define the projector on the $j$-th eigenspace: 
\begin{equation}
    \hat{\Pi}_j = \sum_m \ket{E_{m,j}}\bra{E_{m,j}},
\end{equation}
where $m$ labels the states within the $j$-th  eigenspace. 
Calling $p(n,t)=|\Krylbra{n}\ket{\psi(t)}|^2$ the probability of finding the system in the Krylov state $\ket{K_n}$, one finds
\begin{align}
    \begin{split}
        p(n,t) = \sum_{j,k} \sum_{m,l} &\braket{K_n}{E_{m,j}}\braket{E_{m,j}}{\psi_0} \\ \times&\bra{\psi_0}\ket{E_{l,k}}\bra{E_{l,k}}\ket{K_n}e^{-i(E_j-E_k)t}
    \end{split}
\end{align}
Note that the time-dependent phases depend only on the eigenspace indeces
since $E_{m,j}=E_j$ for every $m$. Hence, the long-time average yields a $\delta_{jk}$, and
\begin{align}
    \begin{split}
       \Bar{p}(n) = \sum_{j=0}^{I-1} |\Krylbra{n}\hat{\Pi}_j\ket{\psi_0}|^2.
    \end{split}
\end{align}
$I$ denotes the number of distinct energy levels of $\hamiltonian$. It follows that the most general expression of $\infavcompl$, holding true for arbitrary spectra, is:
\begin{equation} \label{eq:deginfavcompl}
    \infavcompl = \sum_{n=0}^{\kryldim-1} n \sum_{j=0}^{I-1} |\Krylbra{n}\hat{\Pi}_j\ket{\psi_0}|^2.
\end{equation}
For non degenerate spectra, all projectors are one-dimensional and therefore expression (\ref{eq:deginfavcompl}) reduces to the form \cite{Rabinovici_2022, PhysRevLett.132.160402, Jeevanesan_2024}:
\begin{equation} \label{eq:nondeginfavcompl}
     \infavcompl = \sum_{n=0}^{\kryldim-1} n \sum_{j=0}^{\dimension-1} |\braket{K_n}{E_j}|^2|\braket{E_j}{\psi_0}|^2.
\end{equation}
The form in (\ref{eq:nondeginfavcompl}) can be retrieved even in the degenerate case, by exploiting the nonuniqueness of the energy eigenstates basis. To do so, we consider the \textit{basis of representatives} \cite{Rabinovici_2022, rabinovici2025krylovcomplexity}. 
The key idea is that all eigenstates belonging to the same eigenspace acquire the same phase under time evolution. Consequently, the projection of the initial state onto an eigenspace evolves only by a global phase, so the dynamics explores only the one-dimensional subspace spanned by that projection within each eigenspace of $\hamiltonian$. 
To see this, consider a generic spectral decomposition of the initial state:
\begin{equation}
    \ket{\psi_0} = \sum_{n=0}^{\dimension-1} \psi_n\ket{E_n}.
    \label{eq:spectraldec}
\end{equation}
For illustration purpose, suppose only one energy level of $\hamiltonian$ to be degenerate, and call $\mathcal{E}$ the eigenspace relative to it. Then, (\ref{eq:spectraldec}) becomes 
\begin{equation}
    \ket{\psi_0} = \sum_{n\in \mathcal{E}} \psi_n \ket{E_n} +\sum_{m\notin \mathcal{E}}\psi_m \ket{E_m}.
\end{equation}
The first sum is simply rotated by a global phase through time evolution, so no other directions of $\mathcal{E}$ are explored. Defining the \textit{representative} of $\mathcal{E}$ as the normalized projection of $\ket{\psi_0}$ onto it, 
\begin{equation}
    |E_\mathcal{E}\rangle = \frac{1}{\sqrt{\sum|\psi_n|^2}} \sum_{n\in \mathcal{E}} \psi_n\ket{E_n},
\end{equation}
(\ref{eq:spectraldec}) can be simplified to
\begin{equation}
     \ket{\psi_0} = \mathcal{N}^{-1}\ket{E_\mathcal{E}} + \sum_{m\notin \mathcal{E}}\psi_m \ket{E_m},
\end{equation}
with $\mathcal{N} = \left(\sum|\psi_n|^2\right)^{-\frac{1}{2}}$. Through the basis of representatives, the structure of the Krylov subspace can be fully characterized. Each eigenspace of $\hamiltonian$ that is populated by the initial state contributes with one Krylov dimension. 
Therefore, the dimension $\kryldim$ of the Krylov subspace is equal to the number of eigenspaces of $\hamiltonian$ on which $\ket{\psi_0}$ has nonzero projection \cite{Rabinovici_2022, rabinovici2025krylovcomplexity}. As a corollary, no Krylov subspace can have a dimension exceeding the number of distinct energy levels \cite{Rabinovici_2022, rabinovici2025krylovcomplexity}.
Once the Hamiltonian is diagonalized in the basis of representatives, the matrix elements of the projectors in Eq. (\ref{eq:deginfavcompl}) become 
$\Krylbra{n}\hat{\Pi}_j\ket{\psi_0}=\braket{K_n}{E_j}\braket{E_j}{\psi_0}$, 
with $\ket{E_j}$ the representative associated with the $j$-th eigenspace, 
while $\Krylbra{n}\hat{\Pi}_j\ket{\psi_0}=0$ if $\ket{\psi_0}$ has no component on it. 
Using these observations, the long-time-averaged Krylov complexity can be written as \cite{Craps_2025}:
\begin{equation} \label{eq:minbascompl}
    \infavcompl = \sum_{n=0}^{\kryldim-1} n \sum_{j=0}^{\kryldim-1} |\braket{K_n}{E_j}|^2|\braket{E_j}{K_0}|^2.
\end{equation}
\section{Chiral Quantum walks and quasi-gauge transformations} \label{sec:qwchiral}
Consider a system governed by a Hamiltonian $\hamiltonian$ and evolving according to the evolution operator  $e^{-i\hamiltonian t}$ in a preferred basis $\{\ket{n}\}_{n=0}^{\dimension-1}$. 
The Hamiltonian can be associated with a graph $\graph$ whose nodes correspond to the basis states $\ket n$. Two nodes $j$ and $k$ are connected whenever $\bra j \hamiltonian \ket k \neq 0$, with edge weight given by the corresponding  matrix element. 
The dynamics generated by $\hamiltonian$ on this graph defines a \textit{continuous-time quantum walk}.

A generalization of conventional CTQWs is provided by \emph{chiral quantum walks}, characterized by complex off-diagonal elements in the Hamiltonian,
\begin{align}
H_{jk}=h_{jk}\,e^{i\alpha_{jk}}, 
\label{eq:chiralH}
\end{align}
with $h_{jk},\alpha_{jk}\in\mathbb{R}$.
A particularly important consequence of these phases is that they may lead to the violation of probability time-reversal symmetry (PTS). For a CTQW generated by a real Hamiltonian, Hermiticity implies symmetric transition probabilities, $p_{jk}(t)=p_{kj}(t)$, where $p_{jk(t)}=|\bra{k}e^{iHt}\ket{j}|^2$ is the transition probability between nodes $j$ and $k$ at time $t$.
In contrast, chiral quantum walks can lead to asymmetric transition probabilities, $p_{jk}(t)\neq p_{kj}(t)$, thereby breaking  PTS.
The presence of complex phases, however, is a necessary but not sufficient condition for PTS violation, since some phase configurations can be removed by \emph{quasi-gauge transformations} and therefore have no observable effect on the transport dynamics.

\begin{definition}[Quasi-gauge transformation] \label{def:quasigauge} Let $\{\ket{n}\}_{n=0}^{\dimension-1}$ denote the node  
basis of the graph $\graph$. 
A \emph{quasi-gauge transformation} is a local unitary operator  $\quasigauge $ acting on the nodes as
     \begin{align}
        \quasigauge \ket{n} = e^{i\varphi_n}\ket{n} \quad \forall n
    \end{align}
where $\Vec{\varphi} \in [0,2\pi)^{\dimension}$.
Equivalently, $\quasigauge$ is an element of the group $U(1)^{\dimension}$ acting independently on each node of the graph.
\end{definition}
Under a quasi-gauge transformation, transition amplitudes between nodes acquire only node-dependent phase factors, while the corresponding transition probabilities remain unchanged.
Applying a quasi-gauge transformation to a chiral Hamiltonian allows one to eliminate at most $\dimension-1$  phases from its matrix elements. The total phases accumulated along directed loops of the graph, known as \emph{Aharonov--Bohm phases}, are  quasi-gauge invariants \cite{Turner_2021}. This has important consequences for probability time-reversal symmetry (PTS): tree graphs are always quasi-gauge equivalent to real Hamiltonians, while chiral phases cannot induce PTS breaking in bipartite graphs \cite{PhysRevA.93.042302, Zimbor_s_2013}. Consequently, only non-bipartite chiral graphs can potentially violate PTS.
\subsection{Circulant graphs and Hamiltonians}
\begin{figure}[t]
    \centering
    \begin{subfigure}[t]{0.32\columnwidth}
        \centering
        \includegraphics[width=\linewidth]{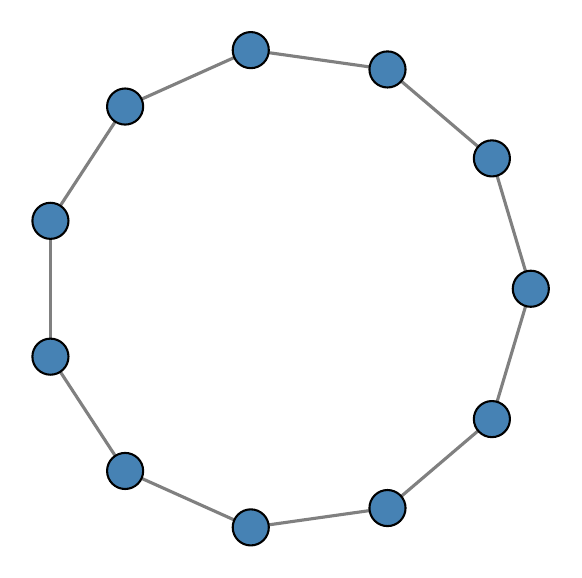}
        \caption{Cycle graph $C_{11}$}
    \end{subfigure}
    \begin{subfigure}[t]{0.32\columnwidth}
        \centering
        \includegraphics[width=\linewidth]{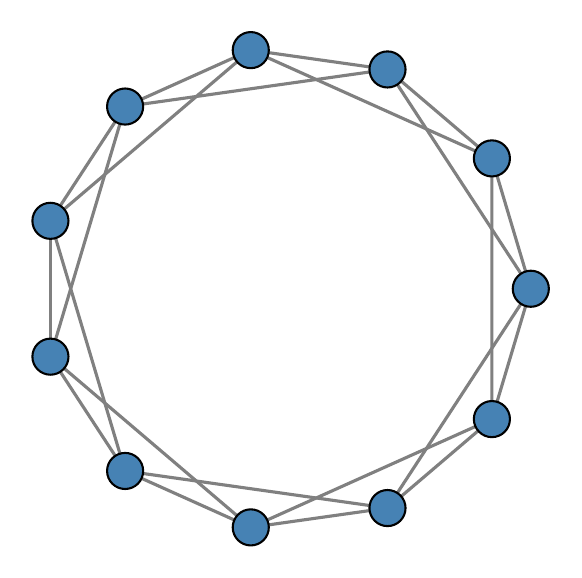}
        \caption{Connectivity $k=2$}
    \end{subfigure}
    \begin{subfigure}[t]{0.32\columnwidth}
        \centering
        \includegraphics[width=\linewidth]{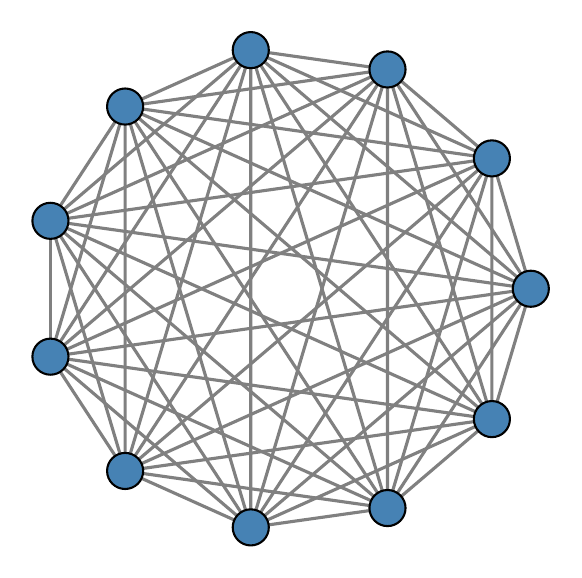}
        \caption{Complete graph $K_{11}$}
    \end{subfigure}
    \caption{Examples of circulant graphs of order $\dimension=11$ with different degrees of connectivity.}
    \label{fig:graphs}
\end{figure}

A graph $\graph$ with $\dimension$ nodes is said to be \emph{circulant} if it is invariant under cyclic translations of the nodes \cite{Olson2014}.
Equivalent statements are: a) its adjacency matrix is circulant, and b) its symmetry group admits the cyclic group $\mathbb{Z}_\dimension$ as a subgroup. 
Cycle graphs and complete graphs are examples of circulant graphs (see Fig. \ref{fig:graphs}) with minimal and maximal connectivity respectively. 
A Hamiltonian defined on a circulant graph does not necessarily inherit the same symmetry, since the edge weights may break translation invariance. In this work, however, we restrict our attention to \emph{circulant Hamiltonians}, namely Hamiltonians that are themselves invariant under cyclic node translations and therefore share the $\mathbb{Z}_{\dimension}$ symmetry of the underlying graph.

The circulant Hamiltonian with connectivity $k$ can be thus  written as 
\begin{equation}
H^{(k)}=\sum_{r=1}^{k}\left(\alpha_rT^r+\alpha^*_rT^{-r}\right),
\end{equation} 
where $T$ is the cyclic translation operator, $T^{\pm r}\ket{n}=\ket{n\pm r \pmod{\dimension}}$, and $\alpha_r$ are complex numbers.
Since $\hamiltonian$ commutes with discrete translations, its eigenstates are the discrete Fourier modes:
\begin{equation} \label{eq:circeigens}
    \ket{E_m} = \cfrac{1}{\sqrt{\dimension}}\sum_{n=0}^{\dimension - 1} e^{\frac{2\pi imn}{\dimension}}\ket{n},
\end{equation}
which form the momentum basis of the graph and can be interpreted as  the discrete analogue of plane waves on a periodic lattice.
While the eigenvectors are fixed by symmetry, the eigenvalues depend on the specific matrix elements of $\hamiltonian$. 
Without loss of generality, we assume the walker to be initially localized at $\ket{0}$. 
All nodes are related by cyclic translations, which are symmetries of the Hamiltonian: as we will prove in Sec. \ref{sec:invariance}, this suffices for the resulting Krylov complexity to be independent of the chosen initial node. 
Expanding the initial state in the Fourier eigenbasis yields
\begin{equation}
    \ket{0} = \frac{1}{\sqrt{\dimension}}\sum_{m=0}^{\dimension - 1} \ket{E_m}.
    \label{0site}
\end{equation}

Therefore, the initial state has equal overlap with all momentum sectors. As a consequence, the Krylov dimension and the long-time-averaged complexity are entirely determined by the degeneracy structure of the spectrum, revealing why circulant Hamiltonians provide a particularly convenient setting for the study of complexity on chiral quantum walks. Their translational symmetry fixes the eigenbasis, so the effect of chiral phases is encoded entirely in the energy spectrum and its degeneracies. This reduction enables an analytical characterization of the long-time-averaged Krylov complexity, which we exploit in the following sections.
\section{$\hamiltonian$-symmetry invariance} \label{sec:invariance}
We  now investigate the behavior of Krylov complexity under the main source of degeneracy, i.e. symmetries. In particular, we  prove that all states belonging to the same orbit of the Hamiltonian's symmetry group  share the same complexity. 
A particular instance of this phenomenon was reported in \cite{PhysRevD.109.L081701}, where Krylov complexity was shown to exhibit a time-translation invariance. 
As we show below, this property is a special case of a broader invariance under unitary symmetries commuting with the Hamiltonian.

\begin{theorem}[Symmetry invariance] \label{thm:complgeneralinv}
    Let $\hat{U}$ be a unitary symmetry operator of the Hamiltonian, i.e. $[\hamiltonian, \hat{U}]=0$, and consider two initial states $\ket{\psi_0}$, $\ket{\psi'_0}$ related by $\hat{U}$, i.e. $\ket{\psi_0'}=\hat{U}\ket{\psi_0}$. 
    Then the complexities of the dynamics seeded by $\ket{\psi_0}$ and $\ket{\psi'_0}$ are equal at all times:
    \begin{align}
        \comploftime{t}'=\comploftime{t}\quad \forall t
    \end{align}
\end{theorem}

\begin{proof}
Consider the polynomial representation of the Krylov basis seeded by $|\psi_0'\rangle$ \cite{balasubramanian2025variationsthemekrylov}:
    \begin{equation}
        |K_n'\rangle = P'_n(\hamiltonian)|\psi_0'\rangle.
    \end{equation}
The polynomial $P'_n$ is primed because, in principle, the Lanczos coefficients generated by $\ket{\psi'_0}$ and $\ket{\psi_0}$ might differ. 
    Since $\hat{U}$ commutes with $\hamiltonian$ and is unitary, the moments of $\hamiltonian$ are not changed by its action:
    \begin{align} \label{eq:invmoments}
        \langle \psi'_0 |\hamiltonian^k|\psi'_0\rangle = \bra{\psi_0}\hamiltonian^k\ket{\psi_0} \quad \forall k.
    \end{align}
    As the Lanczos coefficients are functions of the moments, Eq. (\ref{eq:invmoments}) implies that $(a_n',b_n')=(a_n,b_n)$ as well, for all $n$. Therefore, the polynomials must be $P'_n=P_n$.
    With this property, and again using $[\hamiltonian, \hat{U}]=0$, 
    \begin{equation}
        |K_n'\rangle = \hat{U} P_n(\hamiltonian)|\psi_0\rangle = \hat{U}\Krylket{n}.
    \end{equation}
    In other words,  the Krylov basis of the transformed state is the image under $\hat U$ of the Krylov basis of the original state.
    From this, we compute the amplitudes:
    \begin{align}
        \langle K'_n | e^{-i\hamiltonian t} | K'_0 \rangle &= \Krylbra{n}\hat{U}^\dagger e^{-i\hamiltonian t} \hat{U} \Krylket{0} \\
        &=\Krylbra{n} e^{-i\hamiltonian t} \Krylket{0}.
    \end{align}
    This implies, by inspecting (\ref{eq:compldef}), that $\comploftime{t}$ is symmetry invariant for any $t$.
\end{proof}
The proof relies only on the invariance of the transition amplitudes in the Krylov basis. Consequently, the result immediately extends to all the Krylov moments, such as the Krylov variance of Equation (\ref{eq:krylvariance}) and all higher moments.
Notice how the theorem includes discrete symmetry groups, so the invariance property does not require a continuous symmetry, and hence a Noether charge generating it. Such cases include continuous-time quantum walks, studied in Section \ref{sec:cctqw}, where the symmetries of the underlying graph are  discrete.

\section{Quasi-gauge invariance} 
\label{sec:invprop}
We now demonstrate that quasi-gauge invariance extends to the transfer probabilities on the Krylov chain and, consequently, to the Krylov complexity, whenever the dynamics is initialized in a localized node state.
\begin{theorem}[Quasi-gauge invariance] \label{thm:complinv}
Let $\hamiltonian$ and  $\hamiltonian'$ 
 be two Hamiltonians related by a quasi-gauge transformation $\hamiltonian'=\quasigauge^\dagger\hamiltonian\quasigauge$. 
Let the dynamics be seeded by the same localized initial state, $\Krylket{0}=\ket{j}$.
Then the complexities of the dynamics generated by $\hamiltonian$ and $\hamiltonian'$ are equal at all times:
\begin{align} 
        \comploftime{t}' = \comploftime{t} \quad \forall t.
\end{align}
\end{theorem}

\begin{proof}
Let $\{ \ket{K_n} \}$ and $\{ \ket{K_n'} \}$ denote the Krylov bases generated from the same localized state $\ket{j}$ by $\hamiltonian$ and $\hamiltonian'$, respectively.
Using the polynomial representation of Krylov states, we can write $ \Krylket{n} =  P_n(\hamiltonian)\ket{j}$ and 
 $ \ket{K'_n} =  P'_n(\hamiltonian')\ket{j}$.
Thus, it is sufficient to compare the corresponding Krylov polynomials.
First, the moments are invariant under quasi-gauge transformations: 
\begin{align} 
 \bra{j}\hamiltonian'^k\ket{j} = \bra{j}\quasigauge^\dagger\hamiltonian^k\quasigauge\ket{j} = \bra{j}\hamiltonian^k\ket{j}.
 \end{align}
Since the Lanczos coefficients are determined by the moments, they are identical for $\hamiltonian$ and $\hamiltonian'$.
Hence, $ P'_n = P_n$ and therefore 
\begin{equation}
 P'_n(\hamiltonian')=P_n(\hamiltonian') = \quasigauge^\dagger P_n(\hamiltonian)\quasigauge.
\label{ppn}
\end{equation}
By applying \eqref{ppn}, $e^{-i\hamiltonian't}=\quasigauge^\dagger e^{-i\hamiltonian t}\quasigauge$ and the unitarity of $\quasigauge$, it follows that 
 \begin{align} 
 \label{eq:invtransampl}
\bra{K'_n}&e^{-i\hamiltonian't}\ket{K_0'} = \bra{j}P'_n(\hamiltonian')e^{-i\hamiltonian't}\ket{j} \nonumber\\
&= \bra{j} P_n(\hamiltonian) e^{-i\hamiltonian t}\ket{j} = \bra{K_n}e^{-i\hamiltonian t}\ket{K_0}. 
\end{align}
Therefore the transition amplitudes on the Krylov chain are invariant and, according to Eq. \eqref{eq:compldef} so is the Krylov complexity:
\begin{equation} 
\comploftime{t}'=\comploftime{t}.
\qedhere
\end{equation}
\end{proof}
It is worth noting that Theorem~\ref{thm:complinv} is not a direct consequence of the invariance of inter-node transfer probabilities. Indeed, the Krylov basis vectors are generally delocalized superpositions of node states rather than localized states. The result relies on the stronger fact that quasi-gauge transformations, as just as symmetries, leave the energy distribution of the initial state unchanged and therefore preserve the entire Lanczos construction.


\section{Krylov Complexity of Chiral Quantum Walks} \label{sec:cctqw}
We now investigate how the graph topology and the presence of chiral phases affect the long-time-averaged Krylov complexity. 
%
%
\subsection{Cycle graphs}
We begin with the cycle graph   $C_{\dimension}$.
It can be represented as a regular $\dimension$-gon, whose symmetry group is the dihedral group $D_{\dimension}$, consisting of discrete rotations ($\mathbb{Z}_\dimension$ subgroup) and reflections ($S_2$ subgroup).

For a chiral quantum walk on $C_{\dimension}$, the Hamiltonian can be written, up to a rescaling of time (indexes are  mod $\dimension$) as:
\begin{equation} \label{eq:gencycleham}
    \hamiltonian = \sum_{n=0}^{\dimension-1} e^{i\varphi_{n}}\ket{n+1}\bra{n} +\text{h.c.}
\end{equation}
which contains $\dimension$ phases. As the cycle graph contains a single  loop, there is only one quasi-gauge invariant phase,   $\Phi = \sum_{n=0}^{\dimension-1} \varphi_n$. 
Accordingly, any phase configuration is quasi-gauge equivalent to a circulant Hamiltonian with uniform edge phases $\varphi_n = \frac{1}{\dimension} \Phi$, $\forall\,n$.
By  Theorem \ref{thm:complinv}, any two phase configurations with the same total phase $\Phi$ have identical Krylov complexity, which may therefore be computed in the circulant gauge without loss of generality, provided that the dynamics is initialized in a localized node state.

Let us first analyze the case of   a real  Hamiltonian ($\varphi=0$ in the circulant gauge), whose spectrum is:
\begin{equation} \label{eq:adjspectrum}
    E_m = 2\cos{\left ( \frac{2\pi m}{\dimension} \right )},
\end{equation}
with $m \in \left \{ 0,...,\dimension -1  \right \}$. The reflection symmetry of the Hamiltonian induces a twofold degeneracy $E_m = E_{\dimension-m}$. Hence, eigenstates with opposite momenta $m$ and $\dimension-m\!\!\!\pmod{\dimension}$ belong to the same eigenspace. 
The only non degenerate eigenvalues correspond to $m = 0$ and, for even $\dimension$,  $m=\dimension/2$. 
Thus, the number of distinct eigenspaces of this class of Hamiltonians is:
\begin{align} \label{eq:degeneracycircadj}
    \begin{split}
        &\frac{\dimension + 1}{2} \qquad \text{if $\dimension$ is odd}, \\
        &\frac{\dimension + 2}{2} \qquad \text{if $\dimension$ is even}.
    \end{split}
\end{align}
Since the initial state has nonzero overlap with every eigenspace of $\hamiltonian$, the Krylov dimension coincides with the number of distinct eigenspaces and is therefore given by Eq.~(\ref{eq:degeneracycircadj}).
The corresponding Krylov basis is \cite{Annoni_2024}:
\begin{align}
    \begin{cases}
        \Krylket{0} = \ket{0} & \\
        \Krylket{n} = \frac{1}{\sqrt{2}}\left( \ket{n} + \ket{\dimension-n} \right) & n\neq \dimension-n \\
        \Krylket{\dimension/2} = \ket{\dimension/2} & \text{if $\dimension$ is even}
    \end{cases}.
\end{align}
For odd $\dimension$, the averaged probability of measuring the walker in the $n$-th Krylov state is:
\begin{equation} \label{eq:oddprobs}
    \bar{p}(n) = \frac{2}{\dimension^2}(\dimension-1)
\end{equation}
for $n\neq0$. Then, the infinite-time average of complexity is:
\begin{align} \label{eq:oddcompl}
    \begin{split}
        \infavcompl_{\rm{odd}} &= \frac{1}{4\dimension^2}(\dimension^2-1)(\dimension -1).
    \end{split} 
\end{align}
For even $\dimension$, the contribution of the boundary state $\ket{\dimension/2}$ must be treated separately.
The probabilities for $n\neq 0$ evaluate to:
\begin{align}
        \bar{p}(n\neq\dimension/2) &= \frac{2}{\dimension^2}(\dimension - 2), \\
        \bar{p}(\dimension/2) &= \frac{2}{\dimension^2}(\dimension -1),
\end{align}
yielding:
\begin{align} \label{eq:evencompl}
    \begin{split}
        \infavcompl_{\rm{even}} &= \frac{\dimension}{4}.
    \end{split}
\end{align}
Both Eqs.~\eqref{eq:oddcompl} and \eqref{eq:evencompl} exhibit the same asymptotic scaling, $\infavcompl \sim \frac{\dimension}{4}$, showing that the long-time-averaged complexity grows linearly with system size and remains within the heuristic bounds proposed in \cite{Jeevanesan_2024}.
\newline
\newline
Consider now a nonzero uniform chiral phase $\varphi\in(0,2\pi)$.
The reflection symmetry responsible for the degeneracies of the real spectrum is broken, reducing the symmetry group from $D_{\dimension}$ to $\mathbb{Z}_{\dimension}$.
The  eigenvalues read:
\begin{equation}
    E_m(\varphi) = 2\cos{\left ( \frac{2\pi m}{\dimension} - \varphi \right )}.
\end{equation}
and are generically non-degenerate. The exceptional values $\varphi = k\pi/\dimension$, with $ k\in\mathbb{Z}$ restore the degeneracy structure of the real Hamiltonian. 
Since the corresponding total phase is $\Phi=k\pi$, these cases are quasi-gauge equivalent to real Hamiltonians. By Theorem~\ref{thm:complinv}, they  possess the same Krylov complexity as the real cycle graph and will not be considered further.

For generic $\varphi$, the spectrum is non-degenerate. The basis of representatives therefore coincides with the Fourier modes and the Krylov dimension reaches its maximal value,
$\kryldim=\dimension$.
Using the polynomial representation of Krylov states, 
it follows that
\begin{equation} 
\braket{K_n}{E_m} = P_n(E_m)\braket{K_0}{E_m}. 
\end{equation}
Defining the spectral weights
$p_m = |\braket{E_m}{K_0}|^2$,  Eq.~(\ref{eq:minbascompl}) becomes
\begin{equation} 
\label{eq:complandpols}
        \infavcompl = \sum_{n=0}^{\dimension-1} n \sum_{m=0}^{\dimension-1} p_m^2\,P_n^2(E_m). 
\end{equation}
For a localized initial state on a cycle graph $p_m=\frac{1}{\kryldim}$ for all $m$. Using the normalization of the Krylov states,  $\sum_{m=0}^{\kryldim -1} P^2_n(E_m) = \kryldim$, and that $\kryldim=\dimension$, Eq.~(\ref{eq:complandpols}) reduces to
\begin{equation} \label{eq:maxiprcompl}
    \infavcompl = \frac{1}{2}(\dimension - 1).
\end{equation}
In contrast to the real case, the explicit form of the Krylov basis is not required: the non-degeneracy of the spectrum together with the uniform spectral weights is sufficient to determine $\infavcompl$.
The uniform distribution of spectral weights leading to \eqref{eq:maxiprcompl} implies that the initial state is maximally delocalized in the basis of representatives. Such spread can be quantified by means of the Inverse Participation Ratio (IPR), $\mu[\psi_0]$ on the basis of representatives, defined as:
\begin{equation}\label{eq:ipr} 
\mu[\psi_0] := \sum_{m=0}^{\kryldim-1} |\langle E_m|\psi_0\rangle|^4.
\end{equation}
IPR takes values in $\left[\frac{1}{\kryldim},1\right]$, with minimum and maximal values realized by evenly spread and perfectly localized \footnote{In the basis of representatives, not to be confused with a state localized in a node of the graph.} states respectively. \\
Independently of the system at hand, whenever the initial state has minimal IPR on the representative basis, the complexity of the corresponding dynamics is given by Eq. \eqref{eq:maxiprcompl}, with $\dimension$ in general replaced by $\kryldim$.

These results already suggest that complexity is not a measure of PTS breaking. Indeed, cycle graphs of even length are  bipartite and therefore remain time-reversal symmetric for any chiral phase.
A quantifier of PTS breaking should therefore be insensitive to chirality on such graphs. However, comparing Eq.s~\eqref{eq:evencompl} and (\ref{eq:maxiprcompl}) shows that  complexity changes despite the absence of any PTS breaking.
\begin{figure}
    \centering
    \includegraphics[width=\columnwidth]{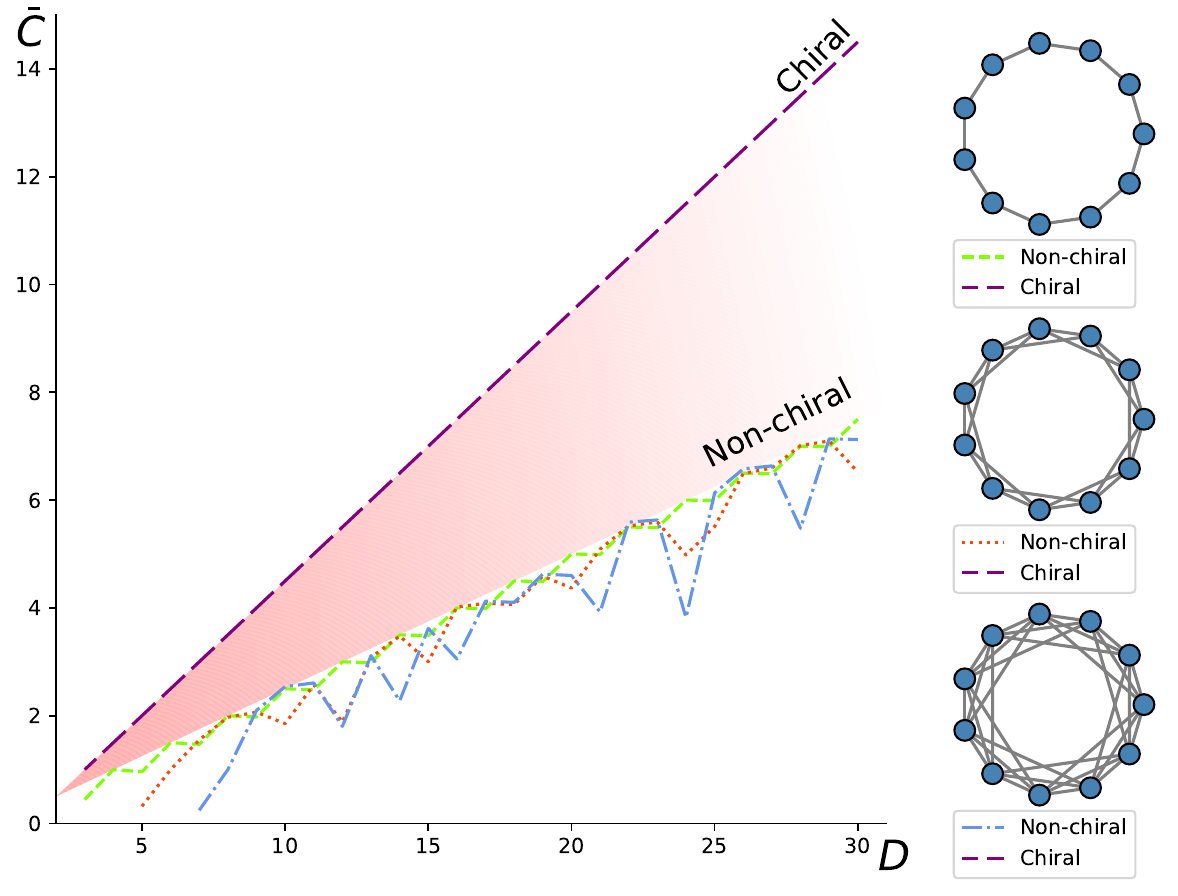}
    \caption{$\infavcompl$ of QWs on circulant graphs seeded by $\ket{0}$. For all topologies studied, the chiral walks explore all the Hilbert space and reach the same value of $\infavcompl=(\dimension - 1)/2$, regardless of the phase values. Real walks, on the other hand, follow topology-specific patterns, with values approximately around $\dimension/4$. }
    \label{fig:circulantgraphs}
\end{figure}

\subsection{Circulant graphs with larger connectivity}
Adding further circulant couplings preserves the cyclic symmetry of the graph but can modify the degeneracy structure of the corresponding real Hamiltonian. As a result, the Krylov dimension and the complexity may change in a nontrivial manner. In the chiral case, instead, the spectrum is generically non-degenerate, so the complexity is determined as in the case of the cycle graph.
The eigenvalues of a circulant Hamiltonian on a graph of $\dimension$ nodes and connectivity degree $k$ read
\begin{equation} \label{eq:circeigenvalues}
    E_m = 2\sum_{r=1}^k \cos{\left(\frac{2\pi rm}{\dimension}-\varphi_r\right )},
\end{equation}
where $\varphi_r$ is defined by $\bra{n+r\pmod{\dimension}}\hamiltonian\ket{n}=e^{i\varphi_r}$, according to Eq. \eqref{eq:chiralH}. \newline
Let us first restrict to real Hamiltonians, namely setting $\varphi_r=0$ for all $r$. For $k>1$, the degeneracy structure of the eigenvalues
is no longer determined solely by the reflection symmetry. Additional accidental degeneracies may arise because different momenta can yield the same value of Eq.~(\ref{eq:circeigenvalues}).
For example, a pentagon graph with $k=2$ coincides with the complete graph $K_5$, whose spectrum contains only two distinct eigenvalues.
For such reason, the long-time-averaged complexity of real Hamiltonians on graphs with $k=2,3$ is computed only numerically (see Appendix \ref{app:compute} for details on the numerical implementation of complexity evaluation). 
The results, displayed in Fig. \ref{fig:circulantgraphs}, confirm the same qualitative scaling $\infavcompl \sim\dimension/4$  found for real cycle graphs. 
\newline
Chiral Hamiltonians, on the other hand, are analytically tractable. 
Consider for instance the $k=2$ graph, with each edge carrying an independent phase. The spectrum depends on the pair of phases $(\varphi_1, \varphi_2) \in (0,2\pi)\times(0,2\pi)$, where $\varphi_1$ and  $\varphi_2$ are associated with first- and second-neighbor edges, respectively. 

Since the Hamiltonian remains circulant, its eigenvectors are still given by the Fourier modes of Eq.~(\ref{eq:circeigens}). Therefore, a localized node state has uniform spectral weights, Eq.~(\ref{0site}), i.e. $p_m=\frac{1}{D}$. 
For generic values of $(\varphi_1,\varphi_2)$, the phases lift the degeneracies of the real spectrum, so that $\kryldim=\dimension$. The same argument used for chiral cycle graphs then applies unchanged, yielding
\begin{equation}
    \infavcompl(\varphi_1, \varphi_2) = \frac{1}{2}(\dimension - 1).
\end{equation}
The same reasoning extends to circulant graphs with arbitrary degrees of connectivity. Whenever the spectrum is non-degenerate, the representative basis coincides with the discrete Fourier modes and a localized node state has uniform spectral weights. It follows that the long-time-averaged complexity is completely determined by the Krylov dimension through Eq.~(\ref{eq:maxiprcompl}).
\newline
\newline
To conclude, let us comment on the discontinuity of $\infavcompl$ at vanishing chiral phases. At first sight, the jump from the real result, $\infavcompl \sim \frac{\dimension}{4}$, to Eq.~(\ref{eq:maxiprcompl}) may suggest that Krylov complexity can resolve arbitrarily small phase differences.
The origin of discontinuity is  that $\infavcompl$ is a  long-time-averaged quantity, i.e. a global property of the dynamics.
As such, it depends on the late-time dynamics, namely the timescale at which infinitesimal differences in the Hamiltonian accumulate and become relevant. 
Finite-time complexity, being a local property, varies continuously with the phases ($\varphi$ is short-hand for an arbitrary number of phases):
\begin{equation}
    \comploftime{t,0} = \lim_{\varphi\rightarrow 0} \comploftime{t,\varphi}. 
\end{equation}
The discontinuity arises only after taking the infinite-time average, because the limits $\varphi\to0$ and $T\to\infty$ do not commute:
\begin{equation}
    \infavcompl(0) = \lim_{T\rightarrow \infty} \lim_{\varphi \rightarrow 0} \frac{1}{T} \int_0^T \comploftime{t}[\varphi]dt \neq \lim_{\varphi\rightarrow 0} \infavcompl(\varphi).
\end{equation}
\subsection{The minimal quantum router}

\begin{figure}[h]
    \centering
    \includegraphics[width=0.6\linewidth]{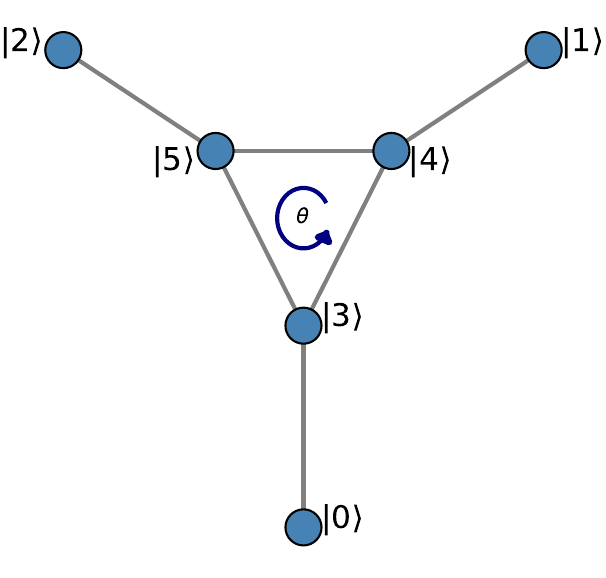}
    \caption{Schematic diagram of the minimal-topology quantum router. 
    The graph only has one loop, thus a single relevant phase circulating in the inner triangle, here called $\theta$. The initial state is the leaf node $\ket{0}$.}
    \label{fig:router}
\end{figure}

\begin{figure}
    \centering
    \includegraphics[width=0.9 \linewidth]{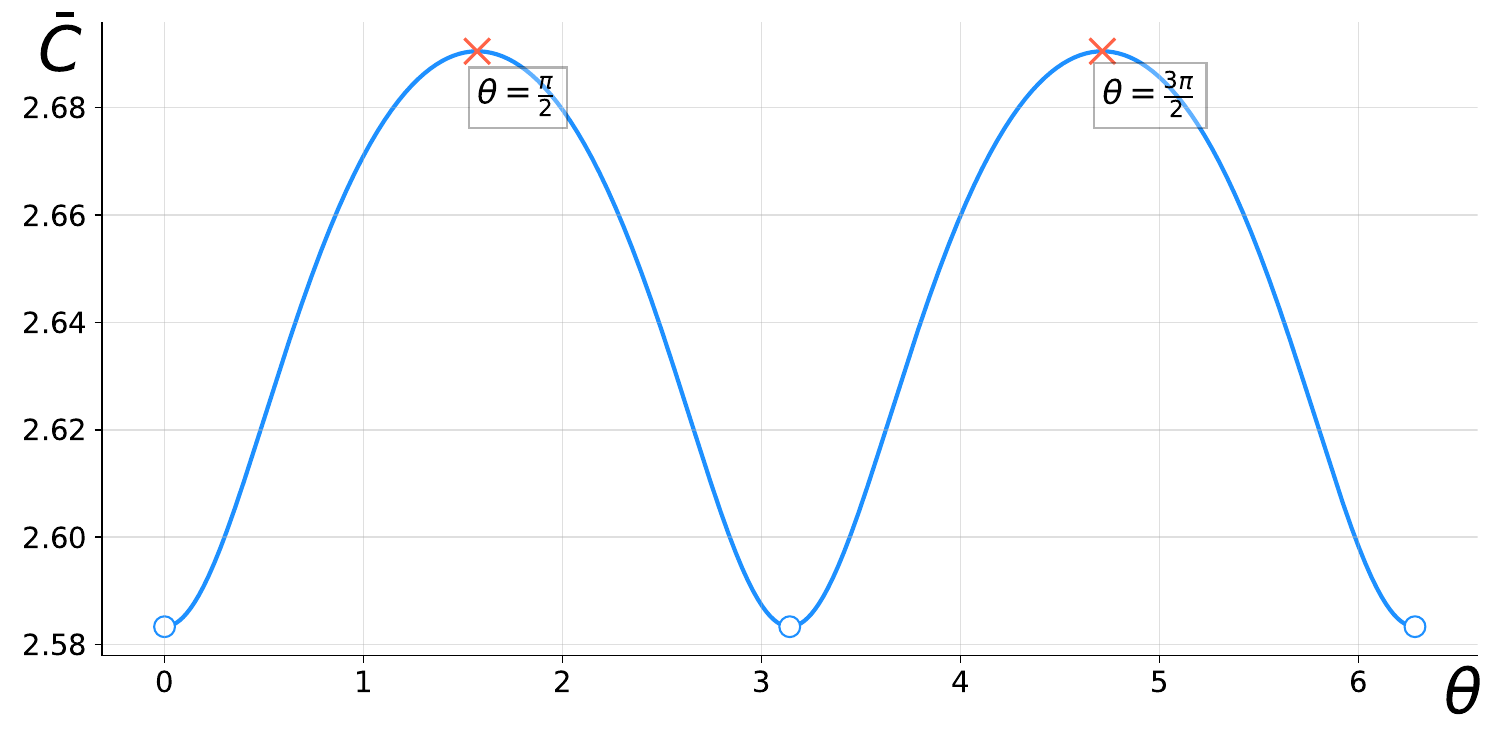}
    \caption{$\infavcompl$ of a quantum router seeded by a leaf node. The empty circles correspond to points of discontinuity, where $\infavcompl=1.3944$. The crosses indicate maximal $\infavcompl$, reached at $\theta \in \left\{\frac{\pi}{2}, \frac{3\pi}{2}\right\}$.}
    \label{fig:routercompl}
\end{figure}

As a final example, let us consider the minimal quantum router displayed in Fig \ref{fig:router}. Following Ref.~\cite{Bottarelli_2023}, we define the associated Hamiltonian 
\begin{align} \label{eq:routerham}
    \hamiltonian(\theta) = \sum_{n=0}^{2} \ket{n+3}\bra{n} + \sum_{n=3}^5 e^{i\frac{\theta}{3}}\ket{n+1}\bra{n} + \text{h.c.}
\end{align}
The graph  exhibits reflection symmetry, inherited by the Hamiltonian only for $\theta = 0, \pi$. 
This is reflected in the spectrum: 
\begin{align} \label{eq:routerspec}
    \begin{split}
        E_n &= \cos{\left ( \frac{\theta}{3} + \frac{4\pi}{3}n  \right )} + \\ &+ (-1)^{n+1}\sqrt{1+\cos^2{\left ( \frac{\theta}{3} + \frac{4\pi}{3}n  \right )}};
    \end{split}
\end{align}
the six distinct energy levels partially coalesce into four at $\theta = 0,\pi$. The eigenstates read
\begin{align} \label{eq:routerstates}
    \ket{E_n} = \frac{1}{\sqrt{3(1+E_n^2)}} \sum_{m=0}^{2} e^{\frac{2\pi i n}{3}m}\left ( \ket{m} + E_n\ket{m+3} \right ).
\end{align}
As opposed to the circulant graphs studied above, the eigenstates of the router depend non-trivially on the phase value. The system is initially localized at $\ket{0}$, which overlaps with all the eigenstates: 
\begin{equation}
    \ket{0} = \frac{1}{\sqrt{3}}\sum_{n=0}^{5} \frac{1}{\sqrt{1+E_n^2}}\ket{E_n}.
\end{equation}
For generic values of $\theta$, the spectrum consists of six distinct eigenvalues and therefore $\kryldim=6$. At the points $\theta\in \{0, \pi\}$, the spectrum becomes degenerate and $\kryldim=4$.
Because of this, we expect once again complexity to be discontinuous around such special values of $\theta$. This is indeed the case, as numerical evaluation of complexity yields $\infavcompl(\theta=0,\pi) = 1.3944$, in contrast with the trend shown in Fig. \ref{fig:routercompl}. 
The maximal complexity is achieved by $\theta \in \left\{\frac{\pi}{2}, \frac{3\pi}{2}\right\}$. These phase values are known to maximise the asymmetry between the transport probabilities $P_{0\rightarrow 1}$ and $P_{0\rightarrow 2}$. 
 
 We also notice that the chiral quantum router is a counterexample to the conjecture that minimal IPR states showcase maximal complexity \cite{PG_2025}. If this was the case, the maximal complexity with a six-dimensional Krylov space would be $\infavcompl_{\text{max}}=(6-1)/2=2.5$, according to Eq. (\ref{eq:maxiprcompl}), in clear contrast with the results obtained.

\section{Discussion} \label{sec:discussion}
In this work, we established general invariance properties of Krylov complexity and applied them to the study of chiral quantum walks.
We proved two invariance theorems for Krylov complexity and, more generally, for all moments of the Krylov distribution under different classes of unitary transformations: symmetries of the Hamiltonian and quasi-gauge transformations. 
Theorem \ref{thm:complgeneralinv} applies to arbitrary initial states and follows solely from the commutation relation between $\hamiltonian$ and its symmetry operators. 
On the other hand, Theorem \ref{thm:complinv} is restricted to localized initial states in the basis diagonalizing the quasi-gauge transformation, because the latter does not in general commute with $\hamiltonian$. 
The common origin of these invariance properties is the preservation of the energy distribution of the initial state. In both cases, the relevant transformation leaves the Hamiltonian moments unchanged and therefore preserves the Lanczos coefficients, implying  identical Krylov   complexity.

We then applied these results to families of chiral quantum walks on circulant graphs. 
Circulant Hamiltonians provide a particularly convenient setting because their eigenvectors are fixed by symmetry, allowing the effects of chirality to be isolated at the level of the spectrum.
We showed that chiral phases generically lift the degeneracies of the corresponding real Hamiltonians, increasing the Krylov dimension and producing a discontinuous change in the long-time-averaged complexity when complex edges are considered. 
This discontinuity originates from the noncommutativity of the limits associated with vanishing phase and infinite averaging time.

Furthermore, for localized initial states, generic chiral circulant Hamiltonians possess a non-degenerate spectrum and uniform spectral weights on the representative basis. As a consequence, their long-time-averaged complexity is completely determined by the Hilbert-space dimension and reaches the universal value $\infavcompl=\frac{1}{2}(\dimension-1)$, independently of the degree of connectivity.
Maximally spread states in the basis of representatives, however, do not realize maximal complexity, as indicated by the example of the quantum router.

Finally, our results indicate that Krylov complexity is not a direct measure of probability time-reversal-symmetry breaking. 
This suggests that Krylov complexity is primarily sensitive to the spectral consequences of chirality, namely the lifting of degeneracies, rather than the breaking of probability time-reversal symmetry itself.

An interesting open question is whether broader classes of transformations exist that preserve the energy distribution of the initial state and therefore leave the associated Lanczos dynamics and Krylov complexity unchanged.

\acknowledgments
The authors thank Simone Cavazzoni, Giovanni Ragazzi, Leo Gentilomo, Gabriele Fazio, Stefano Carini, and Paolo Bordone for useful discussions.
CB is supported by the PRIN 2022 MUR Project EQWALITY (No. 202224BTFZ) and by Unimi through the Piano di Sostegno alla Ricerca  Linea 2 (2025).

\appendix
\section{Computation of Krylov spread complexity} \label{app:compute}
Several implementations of the Lanczos algorithm are prone to numerical instabilities due to mounting rounding errors in the iterative computation of the Lanczos coefficients. This constitutes an issue for high-dimensional problems in particular, as the Krylov states returned by the procedure can be non orthogonal, or even linearly dependent on the previous ones. For this reason, the code used in this work and publicly available on the author's GitHub \cite{complcomputer}, adopts a different approach, first suggested in \cite{Balasubramanian_2022}, based on extracting the Lanczos coefficients by computing the Hessenberg form of the Hamiltonian. The Python library \texttt{SciPy} \cite{10.1145/1141885.1141887} contains implementations of numerically stable algorithms that perform such a task by using Householder reflections instead of iterative orthogonalization procedures, like Lanczos and Gram-Schmidt. Only two details have to be specified. The first is that the implementation assumes the initial state to take the form $(1,0,...,0)^T$. This issue is easily overcome by performing a unitary transformation such that the initial state is the first element of the basis. Such unitary is also performed by use of Householder reflections. Second, it may happen that the first off-diagonal elements are negative, so their absolute values are identified with the $b_n$'s. This just corresponds to multiplying by $-1$ the corresponding Krylov vectors, which does not affect the physics. As we are also interested in evaluating the Krylov subspace dimension $\kryldim$, the obtained matrix is reshaped so as to contain only the first non-zero $b_n$'s. In order to discriminate between null and positive $b_n$'s, a threshold parameter $\epsilon$ settable at run-time is defined in the code. If $b_n\leq \epsilon$, it is considered $0$ and the Hamiltonian is reshaped to its first $n-1$ rows and columns. $\kryldim$ is then computed easily as the number of rows, or equivalently of columns, of the reduced Hamiltonian.

\bibliography{references}

\end{document}